\documentclass[a4paper,fleqn]{cas-sc}

\usepackage[numbers]{natbib}

\usepackage{amsmath}
\usepackage{amssymb}
\usepackage{bm}
\usepackage{color}
\usepackage{commath}
\usepackage{dcolumn}
\usepackage{graphicx}
\usepackage{hyperref}
\usepackage{subfigure}

\begin{document}
\let\WriteBookmarks\relax
\def\floatpagepagefraction{1}
\def\textpagefraction{.001}
\shorttitle{Lateral predictive coding}
\shortauthors{G. Cai et~al.}


\title [mode = title]{Response time of lateral predictive coding and benefits of modular structures}

\author[1]{Guanghui Cai}
\fnmark[1]

\affiliation[1]{
  organization={Institute of Theoretical Physics, Chinese Academy of Sciences},
  addressline={ZhongGuanCun East Road 55},
  city={Beijing 100190},
  country={China}
}

\credit{Investigation, Validation, Writing - review and editing}

\author[2,3]{Zhen-Ye Huang}[orcid=0009-0004-4646-8378]
\fnmark[1]

\affiliation[2]{
  organization={Westlake Institute for Advanced Study},
  city={Hangzhou 310024},
  country={China}
}

\affiliation[3]{
  organization={Center for Interdisciplinary Studies and Department of Physics, School of Science, Westlake University},
  city={Hangzhou 310030},
  country={China}
}

\credit{Investigation, Validation, Writing - review and editing}

\author[1]{Weikang Wang}[orcid=0000-0002-4783-7540]
\cormark[1]
\ead{wangwk@itp.ac.cn}

\credit{Funding acquisition, Supervision, Writing - review and editing}

\author[1,4,5]{Hai-Jun Zhou}[orcid=0000-0003-4228-4438]
\cormark[1]
\ead{zhouhj@itp.ac.cn}

\credit{Conceptualization, Funding acquisition, Investigation, Writing - original draft, Writing - review and editing}

\affiliation[4]{
  organization={School of Physical Sciences, University of Chinese Academy of Sciences},
  city={Beijing 100049},
  country={China}
}

\affiliation[5]{
  organization={Institute for Advanced Physical Studies, Zhejiang University},
  city={Hangzhou 310027},
  country={China}
}

\cortext[cor1]{Corresponding authors.}
\fntext[fn1]{These authors contributed equally.}

\begin{abstract}
  Lateral predictive coding (LPC) is a simple theoretical framework to appreciate feature detection in biological neural circuits. Recent theoretical work [Huang \emph{et al.}, Phys.~Rev.~E \textbf{112}, 034304 (2025)] has successfully constructed optimal LPC networks capable of extracting non-Gaussian hidden input features by imposing the tradeoff between energetic cost and information robustness, but the resulting dynamical systems of recurrent interactions can be very slow in responding to external inputs. We investigate response-time reduction in the present paper. We find that the characteristic response time of the LPC system can be minimized to closely approaching the lower-bound value without compromising the mean predictive error (energetic cost) and the information robustness of signal transmission. We further demonstrate that optimal LPC networks taking a modular structural organization with extensively reduced number of lateral interactions are equally excellent as all-to-all completely connected networks, in terms of feature detection performance, response time, energetic cost and information robustness.
\end{abstract}

\begin{highlights}

  \item Response speed of feature detection in lateral predictive coding

  \item Response time can be minimized together with energetic cost

\item Modular structure does not compromise speed and energy
\end{highlights}


\maketitle


\section{Introduction}

Lateral interactions are abundant in the biological brain. The retina of the visual system is a prominent example neural circuit with rich  inhibitory lateral interactions~\cite{Srinivasan-etal-1982,Kartsaki-2022}. In the primary visual cortex there are both (vertical) lateral interactions within individual cortical columns and (horizontal) lateral interactions between neighboring cortical columns~\cite{Gilbert-etal-1991, Stemmler-etal-1995}. The biological brain has many hierarchical levels, and information processing is carried out in a hierarchical manner through feedforward, feedback, and lateral interactions. Bottom-up information originating from the sensory receptors and top-down message generated by the higher-order associative regions of cerebral cortex are flowing along these hierarchical levels through feedforward and feedback connections~\cite{Lin-Tegmark-2016b,Mikulasch-etal-2022,Millidge-etal-2022,Huang-2022}. The converging feedforward and feedback inputs are processed and integrated within each individual hierarchical level through lateral interactions between peer neurons~\cite{Srinivasan-etal-1982,Bastos-etal-2012}.  Although lateral interactions have not yet been emphasized in current  artificial deep neural networks, neuroscientists believe that they are indispensable for extracting salient features from biological sensory inputs and for reducing the energetic cost of information transmission. 

From the viewpoint of lateral predictive coding (LPC)~\cite{Srinivasan-etal-1982,Harpur-Prager-1996,Ali-etal-2022,Huang-etal-2022,Zhang-etal-2025}, the main biological function of lateral interactions is to filter out the redundancy in the input signals of different peer neurons. When there is statistical correlation between the input signals $s_i$ and $s_j$ of two neurons $i$ and $j$, it may not be necessary for both neurons to respond with high activities $x_i$ and $x_j$; instead neuron $j$ may send an internal signal to neuron $i$ to reduce its net input by an amount $w_{i j} x_j$, and neuron $i$ may also suppress the net input to neuron $j$ by an amount $w_{j i} x_i$. Through experience and learning, the synaptic weights $w_{i j}$ and $w_{j i}$ of these lateral directed interactions are gradually adjusted to best encode the statistical regularity of the input signals, and as a result the magnitudes of the output signals $x_i$ and $x_j$ can be much reduced in comparison to the raw input signals $s_i$ and $s_j$. Because of lateral interactions, the neural activities $x_i$ and $x_j$ follow an internal dynamical process, and the whole system is therefore a recurrent neural network~\cite{Grossberg-2013}.

Lateral predictive coding networks, being an important type of biological computing circuits, have also been studied theoretically in recent years from the perspective of statistical physics. Different from random recurrent neural networks~\cite{Zhang-Fan-etal-2021,Qiu-Huang-2024}, the synaptic weights of the LPC network are far from being random variables; and different from conventional associative memory networks such as the Hopfield model~\cite{Xiong-Zhao-2010,Barra-etal-2012,Sollich-etal-2014}, the synaptic weights of LPC are strongly non-symmetric. To fix all the synaptic weights of an LPC system is essentially an optimization problem under energetic and functional constraints. As the synaptic weight matrix $\bm{W} = \{w_{i j} \}$ adapts to the input signals, some useful collective properties emerge in the LPC network. For example,  we found that simplified models of LPC networks can extensively reduce the statistical correlations among the outputs of individual neurons~\cite{Huang-etal-2022,Huang-etal-2023}. Such a property is consistent with known experimental measurement data on biological neuron networks~\cite{Schneidman-etal-2006}. Model LPC networks can also attain the function of detecting a high-dimensional  non-Gaussian feature which is fully masked by Gaussian background noise, and they can decompose mixed non-Gaussian features into non-orthogonal independent components in the presence of strong background Gaussian noise~\cite{Huang-etal-2025}. 

In our previous theoretical studies we have considered the optimal lateral predictive coding problem from the angle of energy--information robustness tradeoff. The optimal LPC weight matrix $\bm{W}$ is the result of a global optimization process such that  the average strength (energy) of the high-dimensional output signal vector will be minimized and at the same time the information robustness of the output signal will be maximized. The units of a LPC system are strongly competing with each other through lateral interactions, and the system is a recurrent neural network with intrinsic time scales. An important issue not yet addressed is the response speed of the LPC system to input signals. In our previous work we found that when the mean energy decreases, the response times of the LPC dynamical systems increase accordingly, and the systems often become very slow to decay to a steady state. We did not impose any constraint on the total number of synaptic connections, so the optimal synaptic weight matrices were fully connected dense matrices, which are not economic in the sense of wiring cost and internal message-passing.

The possible effects of response speed and number of synaptic connections on the energy--information robustness tradeoff have not yet been explored; they are still open issues in theoretical studies of optimal lateral predictive coding. In the present work, we continue to explore optimal LPC networks based on the same framework of energy--information robustness tradeoff, but emphasizing on the response speed of the network and the wiring cost of synaptic connections. We will demonstrate that decreasing the characteristic response time $\tau_{\textrm{R}}$ of the LPC network will not lead to an increase in the mean energy. And we will also demonstrate that the number of synaptic connections can be extensively reduced by taking a modular structure, without affecting the performance of the network. These numerical results help us to gain deeper understanding on lateral predictive interactions. The present work may also be relevant to the issue of designing energy-efficient artificial recurrent neural circuits.

\section{Model and notations}
\label{sec:model}

We briefly review the lateral predictive coding (LPC) model here and introduce the notations of the key quantities. We consider a network formed by $N$ units with indices $i, j, k, \ldots \in \{1, \ldots, N\}$. The input signal to unit $i$ at time $t$ is denoted as $s_i(t)$, and the output signal from this unit is denoted as $x_i(t)$; the column vectors $\vec{\bm{s}}=(s_1, \ldots, s_N)^\top$ and $\vec{\bm{x}}=(x_1, \ldots, x_N)^\top$ specify the input and internal states of the whole system, and the internal state $\vec{\bm{x}}$ is also interpreted as the output state of the system. There are directed lateral interactions between any two units $i$ and $j$ and these interactions are characterized by an $N\times N$ weight matrix $\bm{W}$; the entry $w_{i j}$ at the $i$-th row and $j$-th column of $\bm{W}$ quantifies the strength of synaptic interaction from unit $j$ to unit $i$. Lateral interactions are non-symmetric, so in general $w_{i j} \neq w_{j i}$; self-interactions are not allowed, so all the diagonal elements are set to zero ($w_{i i} = 0$).

\subsection{Characteristic response time}

Given a weight matrix $\bm{W}$, the recurrent dynamics of the LPC system in response to a time-dependent input $\vec{\bm{s}}(t)$ is
\begin{equation}
\frac{\textrm{d} \vec{\bm{x}}(t)}{\textrm{d} t} \, = \,
\vec{\bm{s}}(t) -  \vec{\bm{x}}(t) - \bm{W} \vec{\bm{x}}(t) \; .
\label{eq:lpc}
\end{equation}
Here we have assumed that lateral interactions are linear, and we interpret the summed lateral contribution $\sum_{j\neq i} w_{i j} x_j(t)$ as the internal predictive message received by unit $i$ at time $t$~\cite{Huang-etal-2022}. The temporal trajectory of the internal state $\vec{\bm{x}}(t)$ is a linear response integral,
\begin{equation}
\vec{\bm{x}}(t) \, = \, \int_0^\infty e^{- (\bm{I} + \bm{W}) \tau}\, \vec{\bm{s}}(t- \tau )\, \textrm{d} \tau \; .
\label{eq:lrs}
\end{equation}

Let us denote the $j$-th eigenvalue of the weight matrix by $\lambda_j  = r_j + i \omega_j$, with $r_j$ and $\omega_j$ being the real and imaginary parts. We can perform an eigen-decomposition
\begin{equation}
  \bm{W} \, = \, \bm{U}
  \begin{bmatrix}
    \lambda_1 &  & \\
    & \ddots &  \\
    &  & \lambda_N
  \end{bmatrix}
  \bm{U}^{-1} \; ,
  \label{eq:Weigen}
\end{equation}
with $\bm{U}$ being the complex-valued eigenvector matrix, and then we derive from Eq.~(\ref{eq:lrs}) that
\begin{equation}
\bigl[ \bm{U}^{-1} \vec{\bm{x}}(t) \bigr]_j \, = \, \int_0^\infty e^{-(1+r_j) \tau -i  \omega_j \tau} 
\, \bigl[\bm{U}^{-1}\vec{\bm{s}}(t-\tau)\bigr]_j\,  \textrm{d} \tau \; .
\label{eq:hatxrelax}
\end{equation}
This expression reveals that the temporal trajectory of the  $j$-th element of the transformed internal state vector $\bm{U}^{-1} \vec{\bm{x}}$ depends only on the temporal trajectory of the $j$-th element of the transformed input state vector $\bm{U}^{-1} \vec{\bm{s}}$, and the characteristic decay time is $1/(1+r_j)$ and the time scale of oscillatory behavior is $1/\omega_j$~\cite{Huang-etal-2022}. When the $r_j$ values are different for different dynamical modes, there will be many different characteristic decay times. In the present work we measure the response speed of the network by the longest value of these decay times.  Given an input signal vector $\vec{\bm{s}}$ at time $t= 0$, the characteristic response time of the whole system is then
\begin{equation}
\tau_{\textrm{R}} \, = \, \frac{1}{1 + r_{\textrm{min}}} \; , 
\end{equation}
where $r_{\textrm{min}}$ is the minimum  among all the $r_j$ values,
\begin{equation}
r_{\textrm{min}} \,= \, \min\limits_{l} \bigl[ r_l \bigr] \; .
\end{equation}
In later discussions we will refer to $r_{\textrm{min}}$ as the minimum eigenvalue-real. 

Obviously, $r_{\textrm{min}}$ must be greater than $-1$ for the dynamical process (\ref{eq:lpc}) to be convergent. It is desirable to have a small characteristic  response time $\tau_{\textrm{R}}$ so that the internal state $\vec{\bm{x}}(t)$ can catch up with the input signal $\vec{\bm{s}}(t)$ swiftly. This means that $r_{\textrm{min}}$ should be as large as possible. On the other hand, because all the diagonal elements of $\bm{W}$ are zero, we have $\sum_l r_l = 0$, therefore the minimum value must be non-positive, 
\begin{equation}
r_{\textrm{min}} \, \leq \, 0 \; .
\label{eq:rminmax}
\end{equation}
This means that the response time has a lower-bound value which is unity,
\begin{equation}
\tau_{\textrm{R}} \, \geq \, 1 \; .
\label{eq:tauRmin}
\end{equation}

A major contribution of the present work is to demonstrate that, even under very severe constraints on $\bm{W}$, this lower-bound characteristic response time can be closely approached.

\subsection{Energy minimization under information-robustness constraint}

Given an input signal vector $\vec{\bm{s}}$,  the steady-state output vector $\vec{\bm{x}}$ is
\begin{equation}
    \vec{\bm{x}} \, = \, \frac{\bm{I}}{\bm{I}+\bm{W}} \vec{\bm{s}} \; .
    \label{eq:io}
\end{equation}
It follows from this expression that  $\vec{\bm{x}} = \vec{\bm{s}} - \bm{W} \vec{\bm{x}}$, so the steady-state output $\vec{\bm{x}}$ is the difference between the input signal $\vec{\bm{s}}$ and the internal prediction vector $\bm{W} \vec{\bm{x}}$, and it therefore serves as a prediction-error vector~\cite{Srinivasan-etal-1982,Huang-etal-2022}. We assume that the energetic cost of transmitting this prediction-error vector is equal to $\sum_l |  x_l |$, as in our earlier work~\cite{Huang-etal-2025}. This is an $L_1$-norm cost function; it encourages all the units $l$ to have smaller response magnitudes $|x_l|$. Denote by $p_{\textrm{in}}(\vec{\bm{s}})$ the probability distribution of input vectors $\vec{\bm{s}}$, then the mean energetic cost $E$ of the LPC system is
\begin{equation}
E  \, = \, \sum\limits_{l=1}^N \int   p_{\textrm{in}}(\vec{\bm{s}})\ \biggl|  \sum\limits_{m=1}^N \Bigl( \frac{\bm{I}}{\bm{I}+\bm{W}} \Bigr)_{l m} s_m  \biggr|\ \textrm{d} \vec{\bm{s}} \; .
\end{equation}

One objective of lateral predictive coding is to reduce this mean energetic cost as much as possible by encoding the statistical regularity of $p_{\textrm{in}}(\vec{\bm{s}})$ into the synaptic weight matrix. There are $N (N-1)$ adjustable parameters $w_{j k}$ for this optimization problem, if no additional structural constraints are imposed on the form of $\bm{W}$.

The steady-state relation (\ref{eq:io}) is a linear mapping from inputs $\vec{\bm{s}}$ to outputs $\vec{\bm{x}}$. Serving as an information transmission channel, it is desirable for the output vectors $\vec{\bm{x}}$ and $\vec{\bm{x}}^\prime$ of two slightly different input vectors $\vec{\bm{s}}$ and $\vec{\bm{s}}^\prime$ to be as widely separated as possible~\cite{Bell-Sejnowski-1995}. Since Eq.~(\ref{eq:io}) projects a tiny volume of the input space to a tiny volume of the output space with a rescaling (Jacobian) factor which is the inverse determinant of $(\bm{I}+\bm{W})$, we can quantify the information robustness of this mapping by the logarithm of the rescaling factor~\cite{Huang-etal-2023}. We call this quantity the (differential) entropy:
\begin{equation}
  S \, =  \,
  - \ln \bigl[ \textrm{det}(\bm{I} + \bm{W} ) \bigr]
  \, =  \,
  - \frac{1}{2} \sum\limits_{l=1}^N  \ln\bigl[ (1+ r_l)^2 + \omega_l^2\bigr]
  \; .
  \label{eq:S}
\end{equation}

Information robustness (entropy) maximization and energetic cost $E$ are conflicting objectives. At a given level of entropy $S$, the mean energy $E$ of any LPC weight matrix $\bm{W}$ can not be arbitrarily small but has a minimum attainable value. In our earlier efforts~\cite{Huang-etal-2023,Huang-etal-2025}, we have studied phase transitions in lateral predictive coding induced by the tradeoff between energy $E$ and entropy $S$. We found that this tradeoff alone (without applying supervised learning) can drive the spontaneous emergence of feature detection function in the optimal LPC networks.

In the present work, we will add the response speed constraint to the LPC system, and will also introduce modular connectivity property to the weight matrix. Our goal is to check whether the objectives of decreasing the response time and reducing the number of synaptic connections can be accomplished together with energy minimization. We will fix the entropy $S$ to some low values, and perform global minimization to get the optimal synaptic weight matrices, following the same stochastic annealing algorithm of Ref.~\cite{Huang-etal-2023}.

\subsection{Feature detection}

To model biologically meaningful sensory data, we assume that the input signal vectors $\vec{\bm{s}}$ are of the following mixed form:
\begin{equation}
  \vec{\bm{s}} \, = \, a_1 \vec{\bm{\phi}}_1 +
  \sum\limits_{j=2}^N b_j \vec{\bm{\phi}}_j \; .
  \label{eq:s1ng}
\end{equation}
Here $\vec{\bm{\phi}}_j = (\phi_j^1, \ldots, \phi_j^N)^\top$ ($j=1,\ldots, N$) are random  $N$-dimensional unit vectors, and they are mutually orthogonal such that $\vec{\bm{\phi}}_j \bm{\cdot} \vec{\bm{\phi}}_k \equiv \sum_{l} \phi_j^l \phi_k^l = 0$ if $j\neq k$ and $=1$ if $j=k$; the coefficient $a_1$ is a non-Gaussian random variable of zero mean and unit variance,  and all the other coefficients $b_j$ are independent Gaussian random variables of zero mean and unit variance, so the covariance matrix of the input vectors $\vec{\bm{s}}$ is an identity matrix free of any information~\cite{Huang-etal-2025}. The first term of Eq.~(\ref{eq:s1ng}) is a random feature $\vec{\bm{\phi}}_1$ with magnitude $|a_1|$, and the second term means a background Gaussian noise vector of mean magnitude $(N-1)^{1/2}$ in the $(N-1)$-dimensional subspace which is orthogonal to $\vec{\bm{\phi}}_1$. The same feature detection problem (\ref{eq:s1ng}) was investigated in Ref.~\cite{Wang-Lu-2019} as an online learning problem, without building an internal model $\bm{W}$.

For the non-Gaussian coefficient $a_1$, a convenient  setting is to assume that it is symmetric with three discrete values, and the corresponding  probability distribution $q(a_1)$ is
\begin{equation}
q( a_1 ) \, = \,
  \left\{
  \begin{array}{ll}
    (1-p_0)/2 \; ,
    & \quad \quad a_1 = 1/ \sqrt{1-p_0} \; , \\
    p_0 \; , & \quad \quad a_1 = 0 \; , \\  
    (1-p_0) / 2 \; , & \quad\quad a_1 = -1 / \sqrt{1-p_0} \; ,
  \end{array}
  \right.
  \label{eqPa1}
\end{equation}
with $p_0$ being the probability of $a_1=0$. Another non-Gaussian distribution is the parameter-free Laplace distribution $q(a_1) = (1/\sqrt{2}) e^{-\sqrt{2} | a_1|}$. We have tested both distributions in our numerical experiments and have obtained results that are quantitatively very similar. To be concrete, we stick to the discrete distribution (\ref{eqPa1}) with fixed  $p_0=0.7$ in the following discussions. 

To measure how the $N$ units respond differently to the non-Gaussian feature $\vec{\bm{\phi}}_1$, we define an order parameter of sensitivity as
\begin{equation}
  Q \, = \, \max\limits_{j} \biggl[  \frac{| \mu_j |}{\sqrt{\sum_{l=1}^N \mu_l^2} } \biggr] \; ,  \quad\quad \textrm{with} \quad
  \mu_j \, = \, \bigl[ (\bm{I}+\bm{W})^{-1} \vec{\bm{\phi}}_1 \bigr]_j \; .
\end{equation}
Since $\mu_j$ is the steady-state projection of the feature vector $\vec{\bm{\phi}}_1$ to unit $j$, if $Q$ is very close to unity, then a single unit of the LPC system must be very sensitive to the feature direction $\vec{\bm{\phi}}_1$~\cite{Huang-etal-2025}.

\section{Reducing response time of fully connected network}
\label{sec:alltoall}

We first consider fully connected LPC networks by allowing all the elements $w_{j k}$ of the synaptic weight matrix $\bm{W}$ with $j\neq k$ to take non-zero values (the diagonal elements are all zero). The system can be represented by a complete directed graph in which there are two edges of opposite directions between any pair of vertices.

\begin{figure*}
  \centering
  \subfigure[]{
    \includegraphics[angle=270,width=0.31\linewidth]{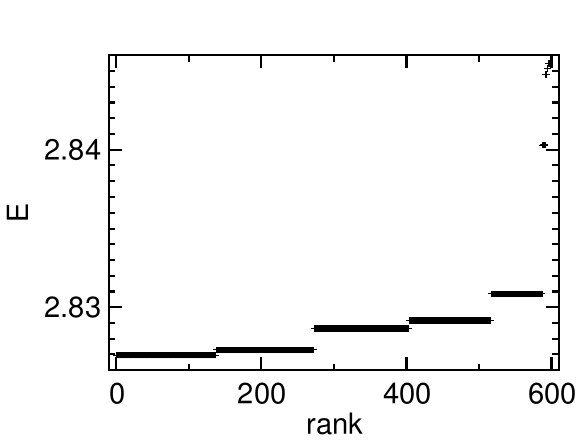}
    \label{fig:lD10Em1em5:E}
  }
  \subfigure[]{
    \includegraphics[angle=270,width=0.31\linewidth]{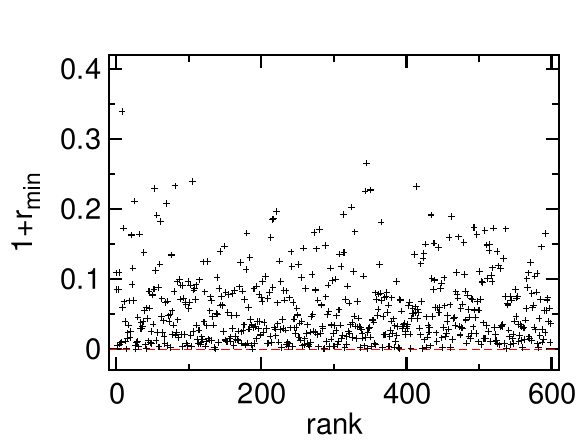}
    \label{fig:lD10Em1em5:r0}
  }
  \subfigure[]{
    \includegraphics[angle=270,width=0.31\linewidth]{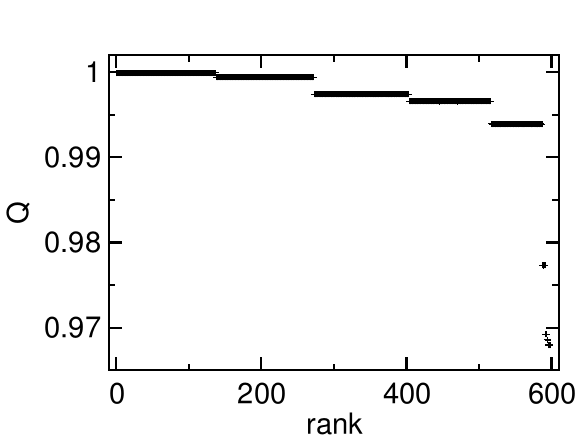}
    \label{fig:lD10Em1em5:Q}
  }
  \\
  \subfigure[]{
    \includegraphics[angle=270,width=0.31\linewidth]{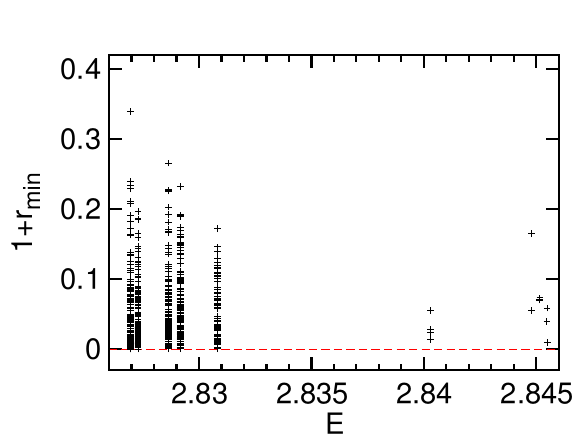}
    \label{fig:lD10Em1em5:Evsr0}
  }
  \subfigure[]{
    \includegraphics[angle=270,width=0.31\linewidth]{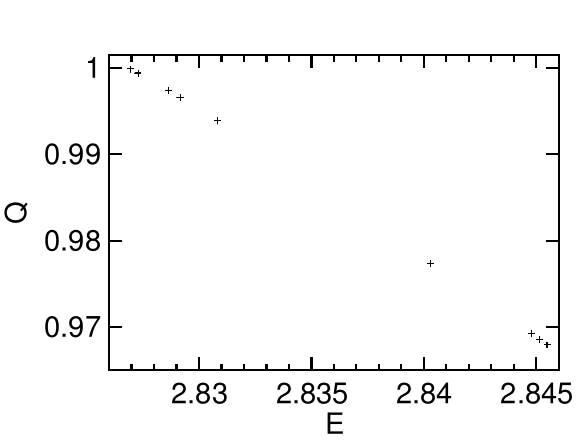}
    \label{fig:lD10Em1em5:EvsQ}
  }
  \caption{(a-c): Simulation results of energy $E$, eigenvalue $(1+r_{\textrm{min}})$, and sensitivity order parameter $Q$ obtained by $600$ independent trials (sorted and ranked in ascending order of $E$) of the stochastic annealing dynamics on a system of size $N=10$ and a fixed random feature direction $\vec{\bm{\phi}}_1$, under very weak constraint of $(1+r_{\textrm{min}}) \geq 10^{-5}$ as marked by the dashed horizontal line of (b). Entropy level is fixed at $S=-10$. (d)-(e): $(1+r_{\textrm{min}})$ versus $E$ and $Q$ versus $E$, for these $600$ trials.}
  \label{fig:lD10Em1em5}
\end{figure*}

\subsection{High degrees of degeneracy of optimal systems}

First, we apply a very weak constraint of $r_{\textrm{min}} \geq -0.99999$ (namely, $1+r_{\textrm{min}} \geq 10^{-5}$) to the weight matrix $\bm{W}$, only to ensure that the dynamics (\ref{eq:lpc}) is convergent.  Figure~\ref{fig:lD10Em1em5} summarizes the numerical results obtained on a relatively small system with $N=10$ units at fixed low entropy value $S=-10$. (Qualitatively similar results are obtained at other low $S$ values such as $S=-5$, $-15$ and $-20$.) For a given random feature direction $\vec{\bm{\phi}}_1$, we repeat the stochastic annealing dynamics independently for $600$ times and then sort the obtained synaptic matrices $\bm{W}$ in ascending order of the mean energy $E$ (Fig.~\ref{fig:lD10Em1em5:E}). We find that $138$ of these $600$ weight matrices have the same minimum energy value $E\approx 2.82695$, and these matrices have considerably  dispersed values of $r_{\textrm{min}}$ (Fig.~\ref{fig:lD10Em1em5:r0} and \ref{fig:lD10Em1em5:Evsr0}), indicating that the optimal weight matrices are highly degenerate. All these optimal matrices are equally excellent at detecting the hidden random feature direction $\vec{\bm{\phi}}_1$, as indicated by the high order parameter $Q \approx 0.9999$ of these optimal matrices (Fig.~\ref{fig:lD10Em1em5:Q} and ~\ref{fig:lD10Em1em5:EvsQ}).

\begin{figure*}
  \centering
  \subfigure[]{
    \includegraphics[angle=270,width=0.31\linewidth]{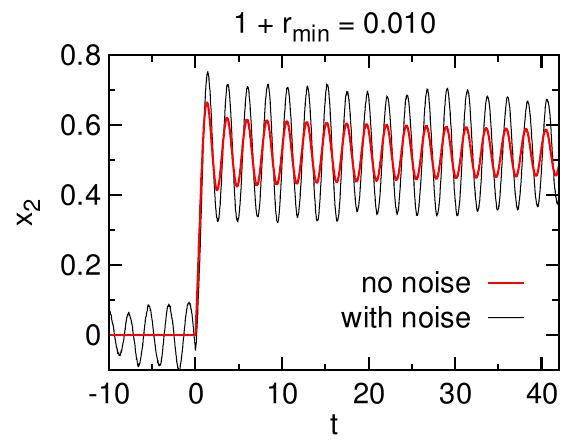}
    \label{fig:lD10Em0p083:x2}
  }
  \subfigure[]{
    \includegraphics[angle=270,width=0.31\linewidth]{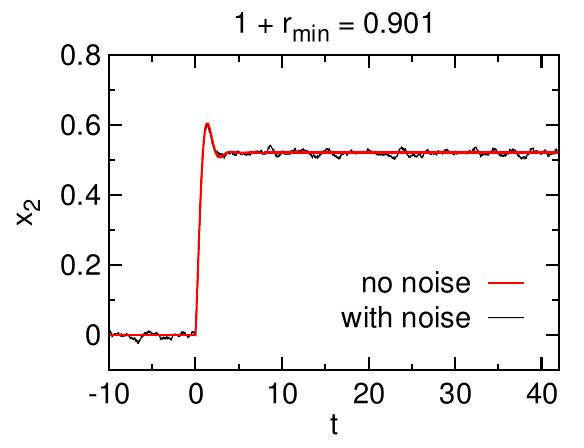}
    \label{fig:lD10Em0p9:x2}
  }
  \caption{
    Some example response trajectories $x_2(t)$ of the most sensitive unit (index $j=2$) to input $\vec{\bm{s}}(t) = a(t) \vec{\bm{\phi}}_1 + \eta \vec{\bm{\varepsilon}}(t)$. The underlying optimal weight matrices $\bm{W}$ are Eq.~(\ref{eq:W010}) with $(1+r_{\textrm{min}}) \approx 0.010$ (a) and Eq.~(\ref{eq:W090}) with $(1+r_{\textrm{min}}) \approx 0.901$ (b). The red thicker lines correspond to pure signal and no noise ($\eta=0$); the thinner black lines correspond to signal plus noise ($\eta=1$). The feature $\vec{\bm{\phi}}_1$ was switched on at $t=0$ with $a(t)$ jumping from $0$ to $1$.
  }
  \label{fig:TrajN10lD10}
\end{figure*}

The value of $r_{\textrm{min}}$ determines the response speed of the predictive coding system. We can clearly demonstrate this by simulating the response behavior of the system to the following test input signals,
\begin{equation}
  \vec{\bm{s}}(t) \, = \,
  a(t) \vec{\bm{\phi}}_1 + \eta \vec{\bm{\varepsilon}}(t) \; ,
  \label{eq:inputst}
\end{equation}
where $\vec{\bm{\varepsilon}}(t) = (\varepsilon_1(t), \ldots, \varepsilon_{N}(t) )^\top$ is the conventional  Gaussian random white noise vector whose elements $\varepsilon_j(t)$ has zero mean and variance $1/N$. We assume that the coefficient $a(t)$ of the feature direction $\vec{\bm{\phi}}_1$ jumps at $t=0$ from $a=0$ (feature absent) to $a=1$ (feature present). To implement Eq.~(\ref{eq:inputst}) numerically, we sample all the elements $\varepsilon_j(t)$ of $\vec{\bm{\varepsilon}}(t)$ independently at each short time step $\delta t = 0.001$ and then update the state vector $\vec{\bm{x}}(t)$ as
\begin{equation}
  \vec{\bm{x}}(t+\delta t) \, = \, e^{- \delta t} \vec{\bm{x}}(t) + ( 1 - e^{ -\delta t } ) \bigl[ \vec{\bm{s}}(t) - \bm{W} \vec{\bm{x}}(t) \bigr] 
  \; .
\end{equation}
Figure~\ref{fig:lD10Em0p083:x2} shows the temporal response $x_2(t)$ of a unit with index $j=2$ (which is the most sensitive unit to $\vec{\bm{\phi}}_1$) at the vicinity of time $t=0$ at zero noise ($\eta = 0$) or at noise comparable to signal ($\eta=1$), obtained on an example optimal network with $r_{\textrm{min}} \approx - 0.990$ (and $1+r_{\textrm{min}}\approx 0.010$). The characteristic response time of this optimal network is  $\tau_{\textrm{R}} \approx 97.1$, which explains the very slow relaxation behavior of $x_2(t)$ in Fig.~\ref{fig:lD10Em0p083:x2}. The dominant oscillatory behavior of $x_2(t)$ is governed by the imaginary part (here $\omega_{\textrm{min}} \approx 2.73$) of the minimum eigenvalue. As $r_{\textrm{min}}$ becomes very close to $-1$, the oscillatory relaxation behavior will extend to hundreds of time units or even longer. Extended oscillation must not be a desirable property for the purpose of rapid response, and it is also energetically unfavorable. 

An encouraging aspect of Fig.~\ref{fig:lD10Em1em5:Evsr0} is that the minimum eigenvalue-real $r_{\textrm{min}}$ at the minimum energy value can take many different values. This aspect is also demonstrated by the highly dispersed low-rank points (rank indices $\leq 138$) of Fig.~\ref{fig:lD10Em1em5:r0}. This phenomenon means that, at the given fixed low value $S$ of entropy, there are many different weight matrices achieving the same minimum energy $E$ while they have quite different response times $\tau_{\textrm{R}}$. In other words, reducing response time $\tau_{\textrm{R}}$ may not necessarily increase the minimum mean energy $E$. Indeed, the partial independence of the response time $\tau_{\textrm{R}}$ with respect to mean energy $E$ can even be rigorously proven for an ensemble of LPC systems with random Gaussian input vectors (see Appendix~\ref{app:igs} for technical details).

\subsection{Optimal systems with short response times}

To search for LPC matrices with short response times $\tau_{\textrm{R}}$, we modify the stochastic annealing algorithm of Ref.~\cite{Huang-etal-2023} by setting a threshold value $r^*$ to the minimum eigenvalue-real $r_{\textrm{min}}$ of the synaptic weight matrix. At each elementary step of the stochastic annealing dynamics, after a change to one row or one column of $\bm{W}$ under the constraint of fixed entropy value $S$ is proposed, we first check the eigen-spectrum of the updated weight matrix: if its $r_{\textrm{min}}$ is lower than the threshold value $r^*$, the proposed update to $\bm{W}$ will be rejected; if $r_{\textrm{min}} \geq r^*$, we then compute the change $\Delta E$ in the mean energy and accept the updated $\bm{W}$ definitely if $\Delta E \leq 0$ or with probability $e^{-\beta \Delta E}$ if $\Delta E >0$. The control parameter $\beta$ is slowly increasing during the simulated stochastic dynamics, following the same protocol  as described in Ref.~\cite{Huang-etal-2023}. 

For the example case of size $N=10$ at fixed entropy $S=-10$, we set several increasingly higher threshold values $r^* = -0.9, -0.8, \ldots, -0.1$; and for each value of $r^*$ we repeat the modified stochastic annealing dynamics $600$ times. At each of these $r^*$ values we find that our annealing dynamics reaches the same minimum energy value multiple times.

\begin{figure*}
  \centering
  \subfigure[$\ (1+r_{\textrm{min}}) \geq 0.90$]{
    \includegraphics[angle=270,width=0.31\linewidth]{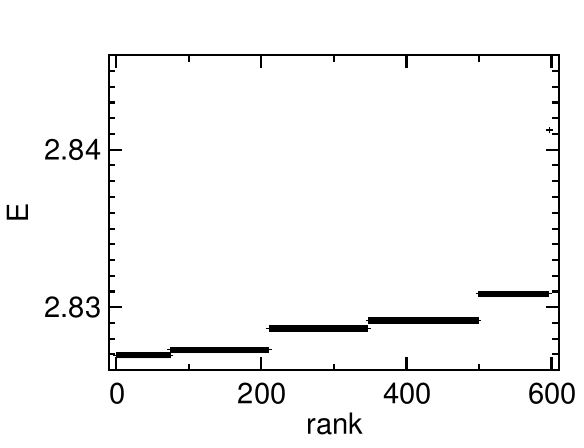}
    \label{fig:lDEm0p9:E}
  }
  \subfigure[]{
    \includegraphics[angle=270,width=0.31\linewidth]{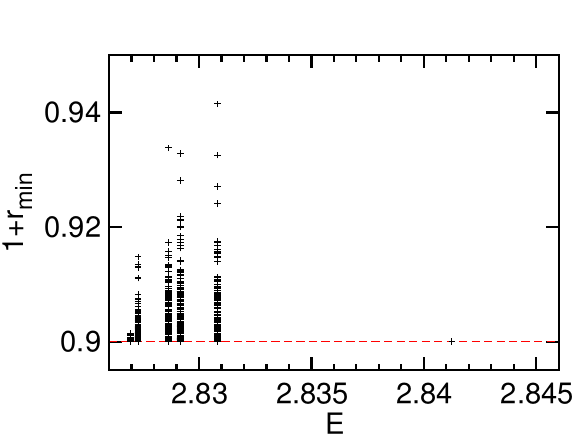}
    \label{fig:lDEm0p9:r0}
  }
  \subfigure[]{
    \includegraphics[angle=270,width=0.31\linewidth]{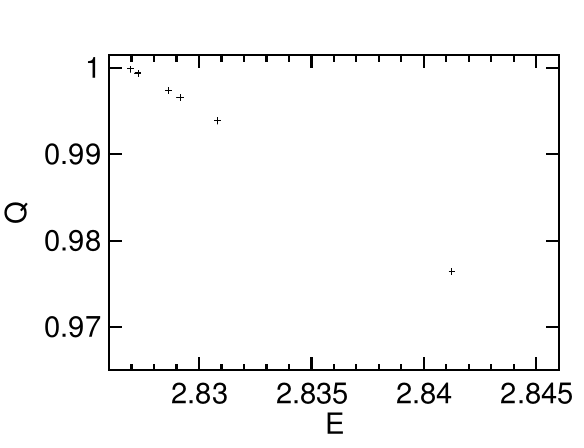}
    \label{fig:lDEm0p9:Q}
  } \\
  \subfigure[$\ (1+r_{\textrm{min}}) \geq 0.99$]{
    \includegraphics[angle=270,width=0.31\linewidth]{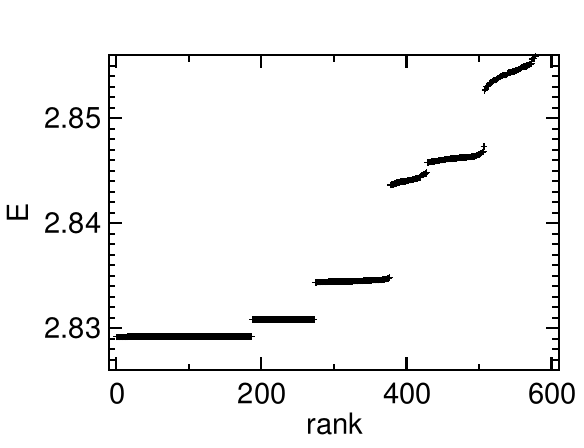}
    \label{fig:lDEm0p99:E}
  }
  \subfigure[]{
    \includegraphics[angle=270,width=0.31\linewidth]{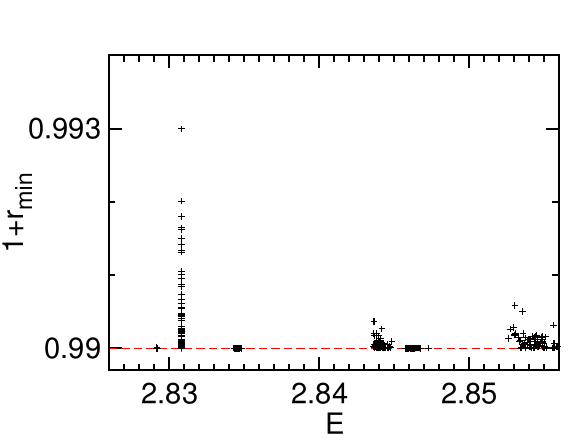}
    \label{fig:lDEm0p99:r0}
  }
  \subfigure[]{
    \includegraphics[angle=270,width=0.31\linewidth]{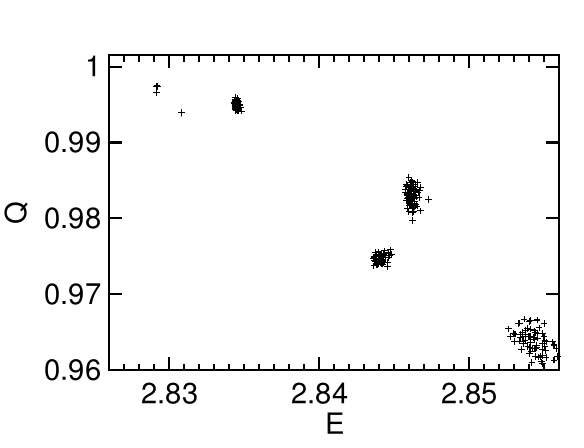}
    \label{fig:lDEm0p99:Q}
  }
  \caption{
    The same system of Fig.~\ref{fig:lD10Em1em5}, but with strong constraint $(1+r_{\textrm{min}}) \geq 0.90$ (a-c), or very strong constraint $(1+r_{\textrm{min}}) \geq 0.99$ (d-f), as marked by the dashed horizontal line of (b) and (e), respectively. (a) and (d): Energy $E$ of the $600$ independent trials (sorted and ranked in ascending order). (b) and (e): Eigenvalue $(1+r_{\textrm{min}})$ versus $E$. (c) and (f): Sensitivity $Q$ versus $E$.
  }
  \label{fig:lDEm0p9}
\end{figure*}

As a concrete example, we show the numerical results obtained at $r^*= - 0.10$ ($1+r_{\textrm{min}}\geq 0.90$, $\tau_{\textrm{R}} < 1.112$) in Fig.~\ref{fig:lDEm0p9} (upper row). Among the $600$ independent trials, $75$ of them reach the minimum energy $E \approx 2.82695$ (Fig.~\ref{fig:lDEm0p9:E}) and their  $r_{\textrm{min}}$ values range from $-0.1000$ to $-0.0987$ (Fig.~\ref{fig:lDEm0p9:r0}) and they are equally excellent at distinguishing the random feature direction $\vec{\bm{\phi}}_1$ with sensitivity $Q \geq 0.9999$ (Fig.~\ref{fig:lDEm0p9:Q}). These optimal networks are much faster at responding to the external input. A typical temporal response trajectory is shown in Fig.~\ref{fig:lD10Em0p9:x2} for an optimal network with $1+r_{\textrm{min}}\approx 0.901$ (and $\tau_{\textrm{R}} \approx 1.11$). We see that, as the feature $\vec{\bm{\phi}}_1$ switches on at $t=0$, the selectively responding unit ($x_2$) changes to its new steady-state value in less than five time units with much suppressed oscillations. 

To see which synaptic interactions are contributing to the huge difference in response time $\tau_{\textrm{R}}$, let us we compare the optimal weight matrix with $(1+r_{\textrm{min}})=0.010$ underlying Fig.~\ref{fig:lD10Em0p083:x2}, 
\begin{equation}
  \bm{W}  = 
  \begin{bmatrix}
    0 & -0.343 &  0.163 & 0.388 & -1.653 & -0.094 & -1.058 & 0.555 & 0.815 & 1.313 \\
    0.330 & 0 & -0.641 & 1.188 & 0.965 & -1.134 & -0.898 & -0.209 & 0.822 & 0.046 \\
    0.046 & 0.777 & 0 & -0.090 & -1.090 & -0.276 & 0.345 & 1.208 & 0.472 & -1.621 \\
    -1.355 & -0.224 & 0.172 & 0 & -0.003 & 1.614 & -0.704 & -0.312 & 1.248 &  -0.508 \\
    1.189 & -0.895 & 0.788 & 0.858 & 0 & -0.271 & 1.094 & 0.222 & 1.032 & -0.365 \\
    0.923 & 0.233 & -0.552 & -1.791 & 0.866 & 0 & -0.524 & 0.524 & 1.091 & -0.134 \\
    1.002 & 0.217 & -1.526 & 0.435 & -1.059 & 0.652 & 0 & -1.139 & 0.182 & -0.697 \\
    -1.183 & 0.114 & -0.977 & -0.156 & -0.150 & -0.413 & 1.486 & 0 & 1.058 & 1.030 \\
    -0.605 & -0.659 & -0.211 & -1.009 &  -0.465 & -1.514 & -0.487 & -1.101 & 0 & -0.823 \\
    0.036 & 0.745 & 1.508 & -0.496 & -0.127 & 0.216 & 0.728 & -1.498 & 0.675 & 0 
  \end{bmatrix}
  \; , 
  \label{eq:W010}
\end{equation}
with the one with $(1+r_{\textrm{min}})=0.901$ underlying Fig.~\ref{fig:lD10Em0p9:x2}, 
\begin{equation}
  \bm{W}  = 
  \begin{bmatrix}
    0 & -0.343 & -1.203 & 0.708 & 0.664 & 0.203 & 1.337 & -0.753 & 0.308 & 1.363  \\
    0.401 & 0 & -1.265 & 0.246 & 1.186 & -0.354 & -0.533 & 0.301 & 0.942 & -1.030 \\
    1.010 & 0.777 & 0 & -1.542 & 0.490 & -0.777 & 1.031 & -0.265 & -0.537 & 0.199 \\
    -0.514 & -0.224 & 1.304 & 0 & 0.499 & -0.574 & 0.945 & -1.048 & 1.131 &  -1.033 \\
    -0.480 & -0.895 & -0.673 & -0.268 & 0 & -1.101 & 0.740 & 1.020 & -1.017 & -0.836 \\
    -0.385 & 0.233 & 0.953 & 0.391 & 0.966 & 0 & 0.570 & 1.873 & 0.532 & 0.782 \\
    -1.487 & 0.217 & -0.835 & -1.119 & -0.920 & -0.398 & 0 & 0.163 & 1.242 & 0.389 \\
    0.716 & 0.114 & 0.249 & 1.035 & -1.029 & -1.904 & -0.204 & 0 & 0.368 & 0.671 \\
    -0.125 & -0.659 & 0.349 &  -0.929 & 1.179 & -0.705 & -1.413 & -0.351 & 0 & 1.029 \\
    -1.555 & 0.745 & -0.025 & 0.794 & 0.650 & -0.624 & -0.226 & -0.679 & -1.199 & 0
  \end{bmatrix}
  \; .
  \label{eq:W090}
\end{equation}
(Only three decimal places are kept for each $w_{j k}$ to save space.) We find that the second column $w_{* 2}$ of the first matrix is identical to that of the second matrix, but all the other columns (and all the rows) are quite different in these two matrices. This means that unit $j=2$ is playing an identical predictive role in both LPC systems, but the internal interactions within the other $(N-1)$ units are completely reorganized in the two systems and they are affecting the $\vec{\bm{\phi}}_1$-sensitive unit by two different sets of synaptic weights $\{w_{2 *}\}$ with row index $j=2$. These comparative results indicate that, although all the other $(N-1)$ units are not sensitive to the feature direction $\vec{\bm{\phi}}_1$, they can greatly help reducing the response time of the single sensitive unit by adjusting their individual synaptic weights to other units.

We can push $r_{\textrm{min}}$ to the upper-bound value (\ref{eq:rminmax}) by setting a threshold value $r^* = -0.01$. The results of $600$ independent trials of the stochastic annealing dynamics are summarized in Fig.~\ref{fig:lDEm0p9} (lower row). Within these $600$ runs, $188$ reach the final energy value of $E\approx 2.8292$ (Fig.~\ref{fig:lDEm0p99:E}), which is slightly but distinctively higher than the minimum energy value $2.82695$ achieved with $r^* \leq -0.1$; the $r_{\textrm{min}}$ values of these $188$ matrices are all very close to $-0.0100$ (Fig.~\ref{fig:lDEm0p99:r0}), and the feature detection performance is slightly compromised with a reduced sensitivity $Q \approx 0.997$ (Fig.~\ref{fig:lDEm0p99:Q}). These results suggest that, with only a very small energetic penalty the typical response time of the predictive coding network can be reduced to a value $\tau_{\textrm{R}} = 1.01$, which is extremely close to the lower-bound value (\ref{eq:tauRmin}).

For the same system of $N=10$, if we decrease the entropy $S$ the minimum energy $E$ will decrease accordingly. We have numerically confirmed that at each lower value of $S$, imposing a very strong constraint on $r_{\textrm{min}}$ does not affect the energy $E$ and sensitivity $Q$ of the optimal networks. To give some numerical data, we list here the values obtained at $S=-20$:
\begin{equation}
\begin{array}{lll}
E\, = \, 1.03996 \; , \quad & Q \, \geq \,  0.99996 \quad\quad &  (r_{\textrm{min}} \, = \, -0.99) \; , \\
E \, = \, 1.03996 \; , \quad & Q \, \geq \,  0.99996  \quad \quad &    ( r_{\textrm{min}}\, = \, -0.10 ) \; , \\
E\, = \,  1.04021 \; , \quad & Q \, \geq \, 0.9989 \quad \quad & ( r_{\textrm{min}} \, = \, -0.01) \; .
\end{array}
\end{equation}

The qualitative conclusion obtained for the system of size $N=10$ also holds for larger systems. For example, when we double the system size to $N=20$, we obtain that:
\begin{equation}
\begin{array}{lll}
  \textrm{At}  \, S  \, =\,  -20: & &  \\
  \, \quad E  \, = \, 5.76216,  & Q \, \geq \,  0.9985 \quad\quad &  (r_{\textrm{min}} \, =\, -0.99) \; ,  \\
  \, \quad E \, = \, 5.76216, & Q \, \geq \,  0.9985 \quad\quad &  (r_{\textrm{min}} \, = \, -0.10) \; ,  \\
  \\
  \textrm{At} \, S  \, = \, - 40 : & & \\
  \, \quad E  \, = \, 2.11940 , & Q \, \geq \,  0.99999  \quad \quad &    ( r_{\textrm{min}}\, = \, -0.99) \; , \\
  \, \quad  E  \, = \,  2.11940 ,  & Q \, \geq \, 0.99999 \quad \quad & ( r_{\textrm{min}} \, = \, -0.10) \; .
\end{array}
\label{eq:N20S20S40}
\end{equation}
As the system size $N$ increases, the number of adjustable parameters (synaptic weights) increases superlinearly, and therefore there are exceedingly greater degrees of freedom to satisfy the additional requirement of high response speed. On the other hand, we need to compute all the eigenvalues of $\bm{W}$ after each proposed modification in our stochastic algorithm, so we can only work with small sizes (less than $N=100$). This may not be a big issue; we will shortly see in Sect.~\ref{sec:modular} that, the computational difficulty associated with large completely connected systems can be easily circumvented by adopting modular structures. 

Taken together, the results of this subsection demonstrate that response time $\tau_{\textrm{R}}$ can be minimized simultaneously with energy $E$ at low levels of entropy $S$. Notice that requiring $r_{\textrm{min}}\approx 0$ will significantly shift all the eigenvalues of $\bm{W}$, and there must be global and large adjustments of predictive interactions. The invariance of minimum energy $E$ is achieved by these extensive modifications of synaptic weights.

\subsection{Distinguishing between two non-Gaussian features}

We also consider the task of detecting and separating two orthogonal or partially aligned non-Gaussian features $\hat{\bm{\phi}}_1$ and $\hat{\bm{\phi}}_2$, defined by
\begin{equation}
\begin{aligned}
\hat{\bm{\phi}}_1 \, & = \,  \cos(\theta /2 ) \vec{\bm{\phi}}_1 + \sin(\theta /2 ) \vec{\bm{\phi}}_2 \; , \\
\hat{\bm{\phi}}_2 \, & = \,  \cos(\theta /2 ) \vec{\bm{\phi}}_1 -  \sin(\theta /2 ) \vec{\bm{\phi}}_2 \; ,
\end{aligned}
\label{eq:ngfs2}
\end{equation}
with $\vec{\bm{\phi}}_j$ again being mutually orthogonal random unit vectors, and $\theta \in (0, \pi /2]$ being the angle between the two feature directions $\hat{\bm{\phi}}_1$ and $\hat{\bm{\phi}}_2$~\cite{Huang-etal-2025}. If $\theta = \pi / 2$ then $\hat{\bm{\phi}}_1$ and $\hat{\bm{\phi}}_2$ are orthogonal, and otherwise they are partially aligned with each other. In our numerical experiments we fix $\theta = \pi/4$ (very similar results are obtained on orthogonal features with $\theta = \pi/2$).

The optimization task for the LPC problem is to build an internal interaction matrix $\bm{W}$ for input signals $\vec{\bm{s}}$ of the following mixed form:
\begin{equation}
  \vec{\bm{s}} \, = \, a_1 \hat{\bm{\phi}}_1 + a_2 \hat{\bm{\phi}}_2 + \sum\limits_{j=3}^{N} b_j \vec{\bm{\phi}}_j \; ,
  \label{eq:stwof}
\end{equation}
with $a_1$ and $a_2$ being two independent random coefficients following the probability distribution (\ref{eqPa1}), and $b_j$ being again independent Gaussian random coefficients as in Eq.~(\ref{eq:s1ng}).

Our stochastic annealing simulation results confirm that the additional constraint on $r_{\textrm{min}}$ does not affect the achievable minimum energy of the LPC system. For a system of size $N=20$ at fixed entropy value $S = -40$, for example, the minimum energy value among $600$ independent runs is $E=2.0442$ at very weak constraint of $r^* = -0.99999$ (the optimal network has $r_{\textrm{min}} = -0.982$, so $\tau_{\textrm{R}} = 55.56$) and at relatively weak constraint of $r^* = -0.9$ (the optimal network has $r_{\textrm{min}} = -0.884$,  $\tau_{\textrm{R}} = 8.621$); and the same minimum energy $E = 2.0442$ is achieved at strong constraint of $r^* = -0.1$ (the optimal network has $r_{\textrm{min}} = -0.0999$, and much shorter $\tau_{\textrm{R}} = 1.111$).

\begin{figure*}
\centering
\subfigure[$\ \hat{\bm{\phi}}_1 =  0.924 \vec{\bm{\phi}}_1 + 0.383\vec{\bm{\phi}}_2\ $]{
\includegraphics[angle=270,width=0.31\linewidth]{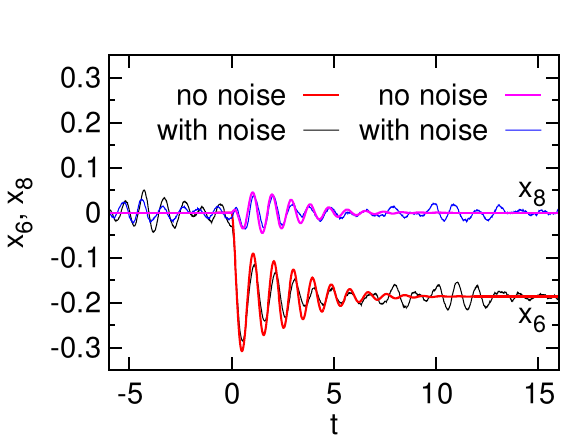}
\label{fig:TFN20lD40E0p9:hatphi1}
}
\subfigure[$\ \hat{\bm{\phi}}_2 = 0.924  \vec{\bm{\phi}}_1 - 0.383 \vec{\bm{\phi}}_2 \ $]{
\includegraphics[angle=270,width=0.31\linewidth]{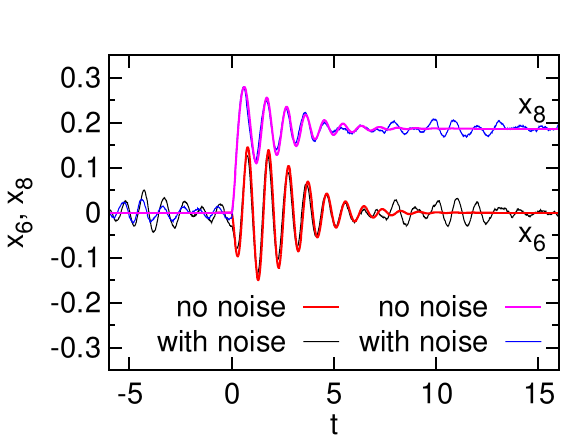}
\label{fig:TFN20lD40E0p9:hatphi2}
}
\subfigure[$\ \vec{\bm{\phi}}_1 = 0.541 (\hat{\bm{\phi}}_1 + \hat{\bm{\phi}}_2) \ $]{
\includegraphics[angle=270,width=0.31\linewidth]{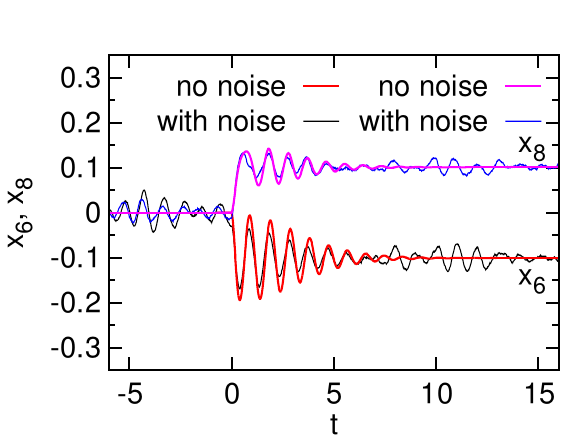}
\label{fig:TFN20lD40E0p9:phi1}
}
\caption{
  Some example response trajectories produced by an optimal LPC network of size $N=20$ and entropy $S=-40$ to the three types of inputs (\ref{eq:threes}). This network has minimum eigenvalue-real  $r_{\textrm{min}} = -0.0999$ and the corresponding imaginary part of eigenvalue is $\omega = 7.3071$; it can distinguish between two non-orthogonal random features $\hat{\bm{\phi}}_1$ and $\hat{\bm{\phi}}_2$ of the form (\ref{eq:ngfs2}) with $\theta = \pi/4$ by the responses $x_6$ and $x_8$ of two units with indices $j=6$ and $k=8$, respectively. The input signal vectors (\ref{eq:threes}) either contain feature  $\hat{\bm{\phi}}_1$ (a), or feature $\hat{\bm{\phi}}_2$ (b), or simply the random feature $\vec{\bm{\phi}}_1$ (c). The red thicker lines correspond to pure signal and no noise ($\eta=0$); the thinner black lines correspond to signal plus noise ($\eta=1$). The feature was switched on at $t=0$ with $a(t)$ jumping from $0$ to $1$.
}
\label{fig:TFN20lD40E0p9}
\end{figure*}

We find that in each of these optimal networks the two feature directions $\hat{\bm{\phi}}_1$ and $\hat{\bm{\phi}}_2$ are captured by two different units with extremely high sensitivity $Q \geq 0.9999$, and each of these two sensitive units is only sensitive to one feature but is indifferent to the other partially aligned feature. We demonstrate this property in Fig.~\ref{fig:TFN20lD40E0p9}, by testing the temporal responses of the above-mentioned optimal network with $\tau_{\textrm{R}} = 1.111$ to three input temporal sequences:
\begin{equation}
\begin{aligned}
\vec{\bm{s}}^{(1)}(t) \, & = \, a(t) \hat{\bm{\phi}}_1 + \eta \vec{\bm{\varepsilon}}(t) \; , \\
\vec{\bm{s}}^{(2)}(t) \, & = \, a(t) \hat{\bm{\phi}}_2 + \eta \vec{\bm{\varepsilon}}(t) \; , \\
\vec{\bm{s}}^{(3)}(t) \, & = \, a(t) \vec{\bm{\phi}}_1 + \eta \vec{\bm{\varepsilon}}(t) \; .
\end{aligned}  
\label{eq:threes}
\end{equation}
The feature is switched on  at time $t=0$ with the coefficient $a(t)$ jumping from $0$ to $1$, and the noise is either absent ($\eta = 0$) or present ($\eta=1$) as in Eq.~(\ref{eq:inputst}).

Under input sequence $\vec{\bm{s}}^{(1)}(t)$ which contains $\hat{\bm{\phi}}_1$, we find that a single unit with index $j=6$ is selectively responding with $x_6 \approx -0.186$ (Fig.~\ref{fig:TFN20lD40E0p9:hatphi1}); under input sequence $\vec{\bm{s}}^{(2)}(t)$ which contains $\hat{\bm{\phi}}_2$, a different unit with index $k=8$ is selectively responding with $x_8 \approx 0.187$ but the state $x_6$ of unit $6$ is silent (Fig.~\ref{fig:TFN20lD40E0p9:hatphi2}). Similarly, the state of unit $8$ is silent ($x_8(t) \approx 0$) if the input sequence is $\vec{\bm{s}}^{(1)}(t)$, see Fig.~\ref{fig:TFN20lD40E0p9:hatphi1}.

On the other hand, if the input sequence is  $\vec{\bm{s}}^{(3)}(t)$ which contains $\vec{\bm{\phi}}_1$, both units $6$ and $8$ are activated but with reduced responses $x_6 \approx -0.101$ and $x_8\approx 0.101$ (Fig.~\ref{fig:TFN20lD40E0p9:phi1}); which means that the system is viewing the random direction $\vec{\bm{\phi}}_1$ as a mixture of two non-orthogonal feature directions, $\vec{\bm{\phi}}_1 \approx 0.541 \hat{\bm{\phi}}_1 + 0.541 \hat{\bm{\phi}}_2$. We have also checked (data not shown) that when the input signal $\vec{\bm{s}}$ contains only $\vec{\bm{\phi}}_2$, the steady-state response of the two units have much higher magnitudes, with $x_6 \approx -0.243$ and $x_8 \approx -0.244$, consisting with the fact that  $\vec{\bm{\phi}}_2 \approx 1.307 \hat{\bm{\phi}}_1 - 1.307 \hat{\bm{\phi}}_2$. We may say that the LPC system is more ``surprised'' with the presence of $\vec{\bm{\phi}}_2$ than with the presence of $\vec{\bm{\phi}}_1$ (this is due to the fact that, at $\theta = \pi/4$, the input signal vectors (\ref{eq:stwof}) are statistically more aligned with the random direction $\vec{\bm{\phi}}_1$ than with $\vec{\bm{\phi}}_2$).

In all the simulated temporal trajectories of Fig.~\ref{fig:TFN20lD40E0p9} we see that the  feature onset-induced oscillations die out quickly in about five time units, which are consistent with the characteristic response time of the underlying network being very short ($\tau_{\textrm{R}}  = 1.111$). 

\section{Response time of modular networks}
\label{sec:modular}

\begin{figure*}
    \centering
    \subfigure[]{
\includegraphics[width=0.22\linewidth]{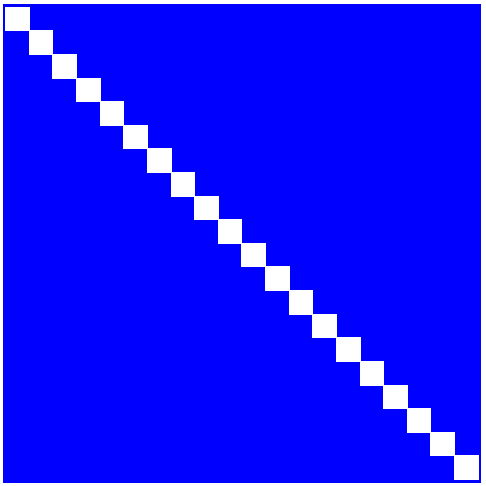}
\label{fig:matrix:single}
}\hspace{0.1\linewidth}
 \subfigure[]{
\includegraphics[width=0.22\linewidth]{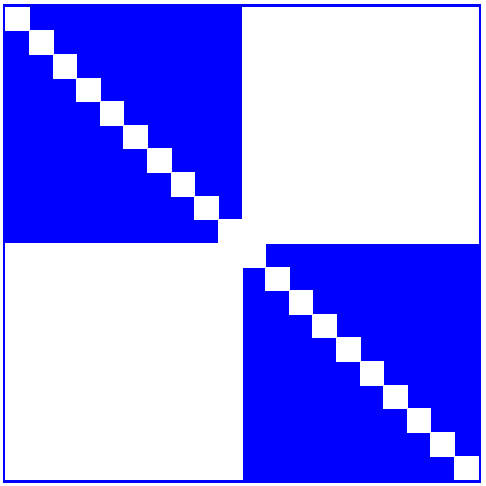}
\label{fig:matrix:indmod}
}\hspace{0.1\linewidth}
 \subfigure[]{
\includegraphics[width=0.22\linewidth]{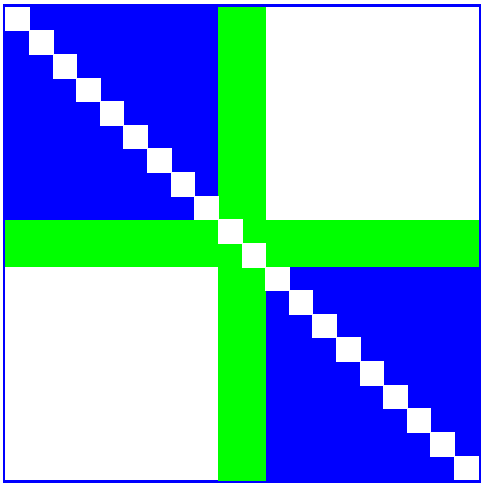}
\label{fig:matrix:ovemod}
}
\\
\subfigure[ single module]{
\includegraphics[angle=270,width=0.31\linewidth]{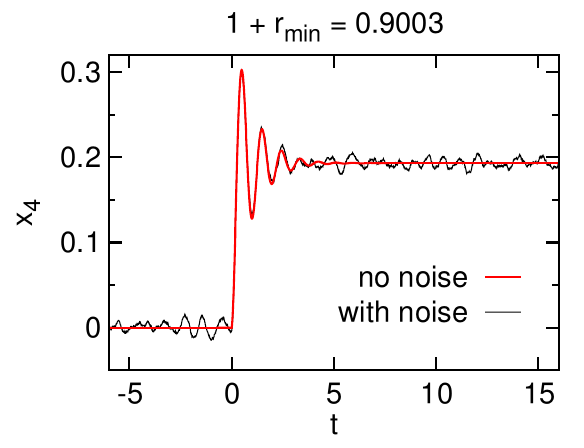}
\label{fig:N20lD40:single}
}
\subfigure[two independent modules]{
\includegraphics[angle=270,width=0.31\linewidth]{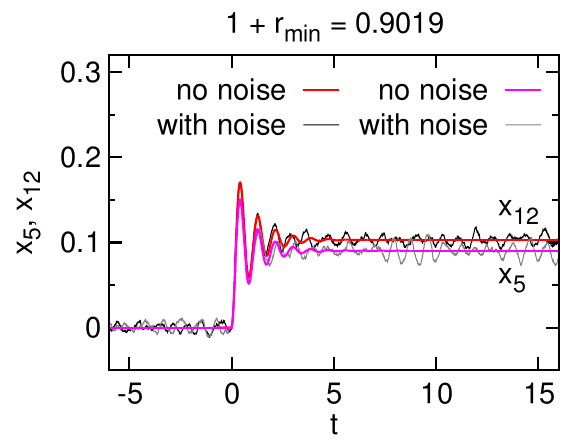}
\label{fig:N20lD40:Block0}
}
\subfigure[two interacting modules]{
\includegraphics[angle=270,width=0.31\linewidth]{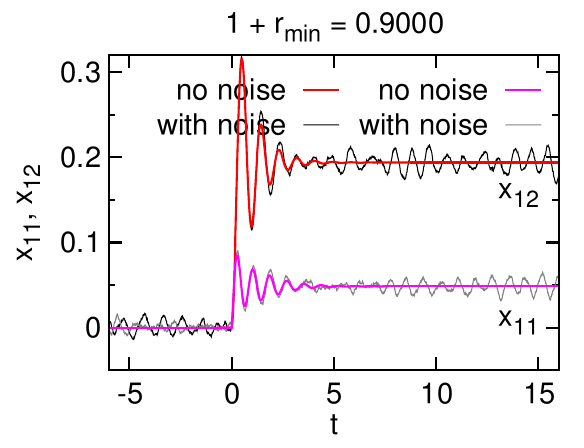}
\label{fig:N20lD40:Blockdb}
}
\caption{
    Illustration of three synaptic weight matrices $\bm{W}$: All-to-all densely connected matrix (a), modular matrix with two mutually independent modules (b), modular matrix with two partially overlapping modules (c). White color indicates zero synaptic weights, blue and green colors indicate non-zero synaptic weights. Number of units is $N=20$. The response properties of these three networks are illustrated in (d)-(f):  A single unit (output state $x_4$) is selectively responding to the random feature $\vec{\bm{\phi}}_1$ in the single-module network (a); two units (output states $x_5$ and $x_{12}$) are selectively responding to $\vec{\bm{\phi}}_1$ in the network with two independent modules (e); two units (output states $x_{11}$ and $x_{12}$) are selectively responding to $\vec{\bm{\phi}}_1$ in the network with two partially overlapping modules (f). The feature $\vec{\bm{\phi}}_1$ is switched on at time $t=0$ with the coefficient $a(t)$ of Eq.~(\ref{eq:inputst}) jumps from $0$ to $1$ at the presence ($\eta=1$) or absence ($\eta=0$) of noise.
}
\label{fig:matrix}
\end{figure*}

The optimal synaptic weight matrices examined in the preceding Sect.~\ref{sec:alltoall} are all densely connected.  There are directed interactions with non-symmetric synaptic weights $w_{j k}$ and $w_{k j}$ between any two units $j$ and $k$; and the total number of synaptic connections is $N (N-1)$, the maximum number of a complete directed graph (see Fig.~\ref{fig:matrix:single}). When the number of units $N$ becomes large, such dense neural networks shall be increasingly undesirable in terms of the various additional costs not included in our simple theory, such as the cost of weights maintenance, the cost of signal transduction within the network, space requirements of the transmission lines and wiring costs~\cite{Liang-etal-2022,Chen-etal-2017,Yu-Yu-2017}. From the purely computational angle,  we have also experienced that as $N$ increases it becomes more and more difficult to properly adjusting the large number of synaptic weights under the severe constraints of entropy $S$ and response time $\tau_{\textrm{R}}$. One would naturally wonder: Are all the synaptic weights really needed to guarantee excellent feature detection and feature separation functions? Will the predictive-coding energy $E$ and characteristic response time $\tau_{\textrm{R}}$ increase greatly if we allow only a small fraction of all the connections to be present in the network?

A fundamental and prominent architectural property of biological neural networks is  modularity~\cite{Salatiello-2026,Sollich-etal-2014,Agliari-etal-2015,Millidge-etal-2024}. The whole neural network can often be decomposed into many smaller structural modules, and the units in individual modules may further divide into sub-modules. The densities of internal connections within the modules are much higher than the densities of connections between different modules~\cite{Zhou-2003-b}. Such a complex hierarchical modular organization is constantly under attack by entropic degradation effects and it can only be stabilized by various biological trade-offs~\cite{Liang-etal-2022,Chen-etal-2017,Yu-Yu-2017}. Here, for simplicity, we will assume that the whole network has already been stabilized in a simple modular structure with several equal-sized modules which are mutually independent (Fig.~\ref{fig:matrix:indmod}) or are partially overlapped (Fig.~\ref{fig:matrix:ovemod}).

\subsection{Matrix with non-interacting modules}
\label{subsec:modnon}

The simplest modular predictive-coding network is a composite network of size $N= k\, n$ formed by $k$ non-interacting modules of size $n$. The whole system can be represented as a collection of $k$ independent complete directed graphs, each of them having $n$ vertices and $n (n-1)$ directed edges, see Fig.~\ref{fig:matrix:indmod}. Given an input signal $\vec{\bm{s}}$ of dimension $N$, each of these modules will process a $n$-dimensional sub-vector free of any predictive interactions from the other modules, and the optimization problem is then equivalent to solving $k$ independent $n \times n$ predictive coding sub-problems. Notice that only a small fraction $(n-1)/(k n-1) \approx 1/k$ of the synaptic interactions of the original single-module complete network are retained in this modular network.

It turns out that greatly reducing the number of synaptic interactions has essentially no adverse effect to the performance of feature detection. Let us demonstrate this fact by a representative example of our numerical experiments.

We divide the predictive coding network of size $N=20$ underlying Eq.~(\ref{eq:N20S20S40}) into $k=2$ non-interacting sub-networks (say, modules $A$ and $B$), each of size $n=10$ (Fig.~\ref{fig:matrix:indmod}). We require the minimum eigenvalue-real $r_{\textrm{min}}$ of both sub-networks to be greater than $r^* = -0.10$ and fix their entropy values both to be $S/2$, with $S=-20$ or $S=-40$ as in Eq.~(\ref{eq:N20S20S40}). When $S = -20$ the mean energy of the optimal modular network is $E=5.862980$, higher by $0.1$ than the value of $5.762160$ of the original single-module network; when $S=-40$, the minimum energy of the modular network is reduced to $E=2.15674$, which is only slightly higher than the corresponding value $2.11940$ of the complete network. These results indicate that reducing the number of synaptic interactions by one-half will only cause negligible increase in the mean energy $E$.

The optimal modular networks at the examined low entropy levels $S$ can all detect the random feature direction $\vec{\bm{\phi}}_1$ of the input signal. Two units of the network, one from each module, gain the function of selectively responding to $\vec{\bm{\phi}}_1$, and their responses are synchronized and equally swift. For example, at $S=-40$, for the randomly generated unit feature vector $\vec{\bm{\phi}}_1$ of our experiment, the unit with index $j=5$ of module $A$ and the unit with index $l=12$ of module $B$ are jointly responding to $\vec{\bm{\phi}}_1$ with high signal-to-noise ratio, and their steady-state activities are $x_5 \approx 0.1031$ and $x_{12} \approx 0.0903$ (Fig.~\ref{fig:N20lD40:Block0}). In comparison, the optimal single-module complete network has only a single sensitive unit (index $i=4$) with respect to $\vec{\bm{\phi}}_1$ and its activity at $S=-40$ is $x_4 \approx 0.1932$ (Fig.~\ref{fig:N20lD40:single}). Notice that the sum $|x_5| + |x_{12}|$ of the two-modules system is equal to $|x_4|$ of the single-module system. This means the two-modules optimal system represents the non-Gaussian feature $\vec{\bm{\phi}}_1$ by two output units but each with output strength reduced by one-half (so the mean energy $E$ remains almost the same). One may argue that an additional advantage of this modular organization is that the presence of the feature $\vec{\bm{\phi}}_1$ can also be inferred by checking the simultaneous activation of $x_5$ and $x_{12}$. 

Although the splitting of the whole LPC system into sub-systems is rather arbitrary in the above-mentioned artificial example, sensory processing modules are well defined and biologically meaningful in real brain systems. In real-world situations, the high-dimensional input signals have intrinsic modular structures and these structures must have been encoded into the LPC systems. That the visual, acoustic, olfactory, tactile features of an object are extracted by different perceptional modalities of the brain in a parallel way and often with comparable response times, may be appreciated as a great achievement of natural selection on the evolutionary time scale.

\subsection{Matrix with interacting modules}

There are many ways to introduce interactions between different modules. For example we may allow the synaptic weights between some randomly picked units of one module and some other randomly picked units of a different module to be non-zero. Here we follow another simple way by assuming that a small number of units are shared by all the different modules.

For the system of size $N=20$, we assume that the units form two overlapping modules, each of size $n=11$, and they share two units (whose indices are assigned as $i=10$ and $i=11$), see Fig.~\ref{fig:matrix:ovemod}. As these two shared units $10$ and $11$ interact with each other and with all the other units of the whole system, they will help achieving mutual predictions between the two modules.

When we solve the same predictive coding problem with a single random feature $\vec{\bm{\phi}}_1$ under the severe entropy constraint of $S = -40$ and severe eigenvalue constraint of $r_{\textrm{min}} \geq -0.10$ (response time $\tau_{\textrm{R}} \leq 1.12$), the resulting optimal solutions achieve a minimum mean energy of $E = 2.12970$, which is lower than the value $2.15674$ of the modular system of the preceding subsection \ref{subsec:modnon} and is very close to the minimum value $2.11940$ of the completely interacting system. We find that a single unit with index $j=12$ from module $B$ is selectively responding to the random feature $\vec{\bm{\phi}}_1$ with high sensitivity $Q \approx 0.9658$, and its activity $x_{12}\approx 0.1940$ in the presence of $\vec{\bm{\phi}}_1$ (Fig.~\ref{fig:N20lD40:Blockdb}) is very similar to the activity value $x_4\approx 0.1932$ of the single responsive unit of the completely connected optimal system (Fig.~\ref{fig:N20lD40:single}). Besides this most responsive output $x_{12}$, one of the two shared units (with index $i=11$) also shows selective response to $\vec{\bm{\phi}}_1$ but its activity $x_{11} \approx 0.0485$ is much weaker (Fig.~\ref{fig:N20lD40:Blockdb}). The other shared unit (index $i=10$) is insensitive to $\vec{\bm{\phi}}_1$. Figure~\ref{fig:N20lD40:Blockdb} also demonstrates that the response time of this modular LPC system is equally excellent as that of the completely connected LPC system.

The numerical results of this subsection confirm the beneficial effect of introducing a few interactions between the different modules. A future extension may be to introduce hierarchical structures into the organization of different modules (see, e.g., Ref.~\cite{Agliari-etal-2015}).

\section{Conclusion and discussions}

In the present work we performed numerical experiments to investigate whether lateral predictive coding networks can minimize their response time $\tau_{\textrm{R}}$ and the mean energy $E$ simultaneously at a given level $S$ of information robustness. For the tasks of detecting a random feature masked by background noise and distinguishing between multiple non-orthogonal random features, we found through global optimization of the synaptic weight matrix that a short response time approaching the theoretical limit is achievable with no additional energetic cost. We further extensively reduced the number of lateral interactions by imposing a modular structure on the predictive coding network, and we demonstrated that such modular networks can achieve the same low levels of response time $\tau_{\textrm{R}}$ and energy $E$ as a complete network.

As an accompanying work to our prior effort~\cite{Huang-etal-2025}, the present paper clarified that short response time $\tau_{\textrm{R}}$ of lateral predictive coding is \emph{not} necessarily a conflicting factor against prediction-error minimization (energetic cost $E$). Our work indicates that lateral predictive coding networks have a great degree of flexibility to accommodate different optimization objectives. The trade-off guiding principle of information robustness versus energetic cost adopted in Refs.~\cite{Huang-etal-2023,Huang-etal-2025} can therefore be extended to the trade-off between information robustness ($S$) on one side and the combined physical factors of energetic cost $E$, response time $\tau_{\textrm{R}}$, and network sparsity on the other side. 

We have not yet worked on very sparse modular LPC networks with total number of directed edges proportional to the total number of units. Another closely related question is to explore sparse and randomly connected networks (without modular structures), with the synaptic weights of the randomly connected edges being adjustable parameters. 

Because LPC matrices $\bm{W}$ are non-symmetric, their eigenvalues $\lambda = r + i \omega$ are complex-valued. When the minimum real part $r_{\textrm{min}}$ of these eigenvalues is maximized toward zero, the eigenvalue imaginary parts $\omega$ are also driven toward certain characteristic values, which are the intrinsic oscillatory frequencies of the LPC system (see Figs.~\ref{fig:TrajN10lD10}, \ref{fig:TFN20lD40E0p9} and ~\ref{fig:N20lD40:single}-\ref{fig:N20lD40:Blockdb} for some concrete examples). Maybe these LPC frequencies are contributing to the intrinsic frequencies of the biological brain. We know that the brain has many frequency bands (such as delta $0.1$-$3$ Hz, theta $4$-$7$ Hz, alpha $8$-$13$ Hz, beta $14$-$25$ Hz, and gamma $25$-$100$ Hz). The complex recurrent interactions of the brain may be conveying predictive messages at various spatial scales. More efforts are certainly needed to study the frequency properties of lateral predictive coding networks.

\section*{Acknowledgments}
The following funding supports are acknowledged: National Natural Science Foundation of China Grants No.~12247104, No.~T2541021  and No.~12447101. Numerical simulations were carried out at the HPC cluster of ITP-CAS and also at the BSCC-A3 platform of the National Supercomputer Center in Beijing. One of the authors (HJZ) thanks the Institute for Advanced Physical Studies of Zhejiang University for hospitality.

\appendix

\section{Some analytical results on the diversity of response times at fixed minimum energy}
\label{app:igs}

Here we present some rigorous analytical results to demonstrate that the response time $\tau_{\textrm{R}}$ of an optimal LPC network can be set to different values without affecting the minimum energy $E$. 

To be analytically tractable, we assume the LPC system is faced with Gaussian input vectors $\vec{\bm{s}}$ of the following form (see Appendix C of Ref.~\cite{Huang-etal-2025}):
\begin{equation}
\vec{\bm{s}} \, = \,  \sqrt{c N} \, a_0 \begin{pmatrix} \frac{1}{\sqrt{N}} \\ \vdots \\ \frac{1}{\sqrt{N}} \end{pmatrix}
+ \sqrt{1 - c} \sum_{k=1}^{N} b_k \vec{\bm{\phi}}_k \; ,
\label{eq:sGs}
\end{equation}
where $c\in [0,1)$ is a constant, and $a_0$ and $b_k$ are independent Gaussian random variables of zero mean and unit variance, and $\vec{\bm{\phi}}_k$ are mutually orthogonal unit vectors as in the main text. The correlation matrix $\bm{C}$ of these Gaussian inputs (\ref{eq:sGs}) is characterized by unity diagonal elements ($C_{k k} = 1$) and constant non-diagonal elements ($C_{j k} = c$ for $j \neq k$). The feature hidden in the inputs $\vec{\bm{s}}$ is then the direction vector $(1, \ldots, 1)^\top / \sqrt{N}$. For such random input vectors $\vec{\bm{s}}$, there exists the following ideal gas law between the minimum energy $E$ and the entropy $S$ of the optimal LPC networks: 
\begin{equation}
\frac{E}{N} \, = \, \sqrt{\frac{2}{\pi}} \bigl( \det(\bm{C}) \bigr)^{\frac{1}{2N}} \exp\Bigl( \frac{S}{N} \Bigr) \; ,
\label{eq:igs}
\end{equation}
where $\det(\bm{C})$ denotes the determinant of $\bm{C}$. This relationship (\ref{eq:igs}) holds if $E/N$ lies between certain upper critical value and certain lower critical value~\cite{Huang-etal-2025}. The optimal weight matrices $\bm{W}$ satisfy the following algebraic condition:
\begin{equation}
    (\bm{I} + \bm{W})(\bm{I} + \bm{W})^{\top} \, = \,  \frac{2}{\pi} T^{-2} \bm{C} \; ,
    \label{eq:Wcod}
\end{equation}
where $T = E/N$ (it is referred to as the temperature in Refs.~\cite{Huang-etal-2025}). 

If the system size is not too small ($N \geq 5$), the solutions for the weight matrix $\bm{W}$ are highly degenerate. For example, one candidate type of solutions is the rotational matrix
\begin{equation}
  \bm{I} + \bm{W} \,  =  \, \begin{bmatrix} 
1 & w_1 & w_2 & \cdots & w_{N-1} \\
w_{N-1} & 1 & w_1 & \cdots & w_{N-2} \\
\vdots & \vdots & \vdots & \ddots & \vdots \\
w_1 & w_2 & w_3 & \cdots & 1 
\end{bmatrix} \; ,
\label{eq:Wrot}
\end{equation}
where $w_j$ are adjustable weight values~\cite{Huang-etal-2023}. Since the correlation matrix $\bm{C}$ has only two eigenvalues: $1 + (N-1)c$, and $(1-c)$ with $(N-1)$ degrees of degeneracy, the complex-valued eigenvalues $(r_k + i \omega_k)$ of $\bm{W}$ must satisfy the following $N$ constraints:
\begin{equation}
\begin{aligned}
 (1+ r_1)^2 + \omega_1^2 \, & = \,  \frac{2}{\pi T^2} \bigl(1 + (N-1) c \bigr) \; , \\
 (1 + r_k )^2 + \omega_k^2 \, & =  \, \frac{2}{\pi T^2} (1-c) \; , \quad\quad \textrm{for}\,\,\, k\geq 2 \; .
\end{aligned}
\label{eq:rjcond}
\end{equation}
In addition, because all the diagonal elements of the weight matrix $\bm{W}$ are identical to zero, we have $\sum_{k=1}^N (r_k + i \omega_k) = 0$. If $N$ is odd, then one of these eigenvalues is real ($\omega_1 = 0$) and the remaining $(N-1)$ of them form complementary pairs; if $N$ is even, then two of these eigenvalues are real and the remaining $(N-2)$ of them form complementary pairs. The $(N-1)$ weight parameters $w_k$ of Eq.~(\ref{eq:Wrot}) are uniquely determined from these $N$ eigenvalues (see supplementary notes of Ref.~\cite{Huang-etal-2023}).
 
The solutions of Eq.~(\ref{eq:rjcond}) are highly degenerate. If $N$ is odd, the solution with the minimum response time $\tau_{\textrm{R}}$ is achieved by
\begin{equation}
\begin{aligned}
r_1 \, & = \,  -1 + \sqrt{\frac{2\bigl(1+(N-1) c\bigr)}{\pi T^2}} \; , \\
r_k \, &  = \, - \frac{r_1}{N-1} \; , \quad \quad (k \geq 2) \; .
\end{aligned}
\label{eq:rmint}
\end{equation}
For example, at correlation $c=0.4$, if the system size is $N=5$ and energy $E=2$ (with $T=0.4$), we have $r_1 = 2.2163$ and $\omega_1=0$, $r_k = -0.5541$ and $|\omega_k| = 1.4793$ for $k\geq 2$, and the minimum response time is $\tau_{\textrm{R}} = 2.2426$. The synaptic weight matrix $\bm{W}$ is
\begin{equation}
\label{examplematrix1}
\begin{bmatrix}
    0 & -0.3565 & 0.7691 & 0.3391 & 1.4647 \\
    1.4647 & 0 & -0.3565 & 0.7691 & 0.3391 \\
    0.3391 & 1.4647 & 0 & -0.3565 & 0.7691 \\
    0.7691 & 0.3391 & 1.4647 & 0& -0.3565 \\
    -0.3565 & 0.7691 & 0.3391 & 1.4647 & 0\\
\end{bmatrix}
\; .
\end{equation}
By adjusting the values of $r_k$ ($k\geq 2$) under the constraints of Eq.~(\ref{eq:rjcond}), we can obtain a continuous spectrum of response times $\tau_{\textrm{R}}$ beyond the minimum value. As a concrete example, for the same system of $N=5$ and $E=2$, if we choose $r_2 = r_5 = -0.9999$ (which yields $\omega_2 = -\omega_5 = 1.5451$), we will get $r_3 = r_4 = -0.1083$ (and $\omega_3 = -\omega_4 = 1.2618$), and a very long response time $\tau_{\textrm{R}} = 10000$. The corresponding weight matrix $\bm{W}$ is
\begin{equation}
\label{examplematrix2}
\begin{bmatrix}
    0 & -0.5297 & 0.8702 & 0.6367 & 1.2392 \\
    1.2392 & 0 & -0.5297 & 0.8702 & 0.6367 \\
    0.6367 & 1.2392 & 0 & -0.5297 & 0.8702 \\
    0.8702 & 0.6367 & 1.2392 & 0& -0.5297 \\
    -0.5297 & 0.8702 & 0.6367 & 1.2392 & 0 \\
\end{bmatrix}
\; .
\end{equation}

The ideal-gas law (\ref{eq:igs}) can also be achieved by other types of LPC networks. To break the rotational symmetry of the matrix (\ref{eq:Wrot}), for example, we may assume the following block structure:
\begin{equation}
  \bm{W} \, = \,  \begin{bmatrix}
    \bm{R} & f \bm{1} \\
    g \bm{1}^\top & 0 
  \end{bmatrix}
  \, = \,
  \begin{bmatrix}
    0 & w_1 & w_2 & \dots & f \\
    w_{N-2} & 0 & w_1 & \dots & f \\
    \vdots & \vdots & \vdots & \ddots & \vdots \\
    g & g & g & \dots & 0 \\
  \end{bmatrix}
  \; , 
  \label{eq:Wrblock}
\end{equation}
where $\bm{R}$ is an $(N-1) \times (N-1)$ rotational matrix whose diagonal elements are all zero,  $\bm{1} = (1, 1, \ldots, 1)^\top$ denotes the $(N-1)$-dimensional unit column vector, and $f$ and $g$ are two additional weight parameters. A concrete example for the same system of $N=5$ and $E=2$ (at $c=0.4$) is
\begin{equation}
  \begin{bmatrix}
    0 & -0.8419 & -0.2933 &  0.0035 &  1.4779\\
    0.0035 &  0& -0.8419 & -0.2933 &  1.4779\\
    -0.2933 &  0.0035 &  0 & -0.8419 &  1.4779\\
    -0.8419 & -0.2933 &  0.0035 &  0 &  1.4779\\
    -0.8630 & -0.8630 & -0.8630 & -0.8630 &  0
  \end{bmatrix}
  \; .
  \label{matrix3}
\end{equation}
The minimal real part of the eigenvalues of this matrix is $r_{\textrm{min}} = -0.5659$, so the response time is $\tau_{\textrm{R}} \approx 2.3035$. One unit (index $j=5$) of this LPC system responds selectively to the mean direction vector $(1,\ldots, 1)^\top$ of the input vectors $\vec{\bm{s}}$ with high sensitivity $Q \approx 0.9610$.

To solve the optimal LPC problem with the property (\ref{eq:Wcod}), we first obtain from Eq.~(\ref{eq:Wrblock}) that
\begin{equation}
  (\bm{I}+\bm{W}) (\bm{I} + \bm{W})^\top \, = \,
  \begin{bmatrix} 
    (\bm{I}+\bm{R})(\bm{I}+\bm{R})^\top + f^2 \bm{1}\bm{1}^\top & (g v + f) \bm{1} \\
    (g v + f)\bm{1}^\top & 1 + (N-1)g 
  \end{bmatrix}
  \; ,
  \label{eq:M_blocks}
\end{equation}
where $v = 1+\sum_{j=1}^{N-2} w_j$, and $\bm{1}\bm{1}^\top$ is the $(N-1)\times (N-1)$ matrix with all entries being unity. Comparing Eq.~(\ref{eq:M_blocks}) with Eq.~(\ref{eq:Wcod}), we can easily write down the equations for the weight values $f$ and $g$, and obtain a set of equations very similar to Eq.~(\ref{eq:rjcond}) for the eigenvalues of the $(N-1)$-dimensional matrix $(\bm{I}+\bm{R})$. These analytical equations describe a continuous set of optimal weight matrices of the form (\ref{eq:Wrblock}), all of them having exactly the same mean energy $E$ but having different response times $\tau_{\textrm{R}}$.

It may be interesting to notice that, the optimal LPC matrices $\bm{W}$ for the feature detection task (\ref{eq:s1ng}) can also be decomposed into two blocks, similar to but more complex than Eq.~(\ref{eq:Wrblock}).  Can the rigorous analysis of this appendix be extended to prove the broad distribution of response times for the feature detection task (\ref{eq:s1ng})? We leave this as an open issue.

\printcredits

\bibliographystyle{cas-model2-names}


\end{document}